\documentclass[epjST]{svjour}
\usepackage{hyperref}
\usepackage[latin1]{inputenc}
\usepackage[T1]{fontenc}
\usepackage{amsmath}
\usepackage{amsfonts}
\usepackage{amssymb}
\usepackage{array}
\usepackage{color}
\usepackage{graphicx}
\usepackage[left=2cm,right=2cm,top=2cm,bottom=2cm]{geometry}
\usepackage{slashbox}
\usepackage{graphics}
\begin{document}
\title{Droplet actuation induced by coalescence: experimental evidences and phenomenological modeling}
\author{Mathieu Sellier\inst{1}\fnmsep\thanks{\email{mathieu.sellier@canterbury.ac.nz}} \and Volker Nock\inst{2} C\'ecile Gaubert\inst{3} \and Claude Verdier\inst{4}}
\institute{Department of Mechanical Engineering, University of Canterbury, Christchurch, New Zealand \and Department of Electrical and Computer Engineering, University of Canterbury, Christchurch, New Zealand \and Ecole Normale Sup\'erieure, Cachan, France \and Laboratoire Interdisciplinaire de Physique, CNRS and Universit\'e Grenoble I, UMR 5588, Saint-Martin d'H\`eres, France}
\abstract{
This paper considers the interaction between two droplets placed on a substrate in immediate vicinity. We show here that when the two droplets are of different fluids and especially when one of the droplet is highly volatile, a wealth of fascinating phenomena can be observed. In particular, the interaction may result in the actuation of the droplet system, i.e. its displacement over a finite length. In order to control this displacement, we consider droplets confined on a hydrophilic stripe created by plasma-treating a PDMS substrate. This controlled actuation opens up unexplored opportunities in the field of microfluidics. In order to explain the observed actuation phenomenon, we propose a simple phenomenological model based on Newton's second law and a simple balance between the driving force arising from surface energy gradients and the viscous resistive force. This simple model is able to reproduce qualitatively and quantitatively the observed droplet dynamics.}
\maketitle
\section{Introduction}
\label{intro}
Microfluidic devices play an ever increasing role in nano- and biotechnologies. An emerging area of research in this technology-driven field is digital microfluidics which is based upon the micromanipulation of discrete droplets. Microfluidic processing is performed on unit-sized packets of fluid which are transported, stored, mixed, reacted, or analyzed in a discrete manner. An obvious challenge however is how to displace the sessile droplets on a substrate. This work investigates a little explored driving mechanism to actuate droplets: the surface energy gradient which arises during the coalescence of two droplets of liquid having different compositions and therefore surface tensions. The resulting surface energy gradient gives rise to a Marangoni flow which, if sufficiently large, can displace the droplet in analogy to thermocapillary actuation which has been extensively studied, see for example the review of Darhuber and Troian, \cite{Darhuber05}, or the recent paper of Gomba and Homsy, \cite{Gomba09}.

One of the earliest documented observations of the displacement induced to a droplet placed in the vicinity of another one dates back to work of Bangham and Saweris, \cite{Bangham38}. The authors described qualitatively the motion of a water droplet on a mica substrate induced by the nearby presence and possible coalescence of a droplet of 60\% acetic acid. They postulated the presence of a very thin film connecting the two droplets. The problem was recently revisited by Bahadur and co-workers, \cite{Bahadur09}, who attempted to unequivocally answer the question of whether the induced motion is the result of the presence of a thin connecting film or transport phenomena through the atmosphere as suggested by Carles and Cazabat, \cite{Carles89}. Their results emphasized the importance of the thin film but could not disregard the role played by the atmosphere since the volatile phase tends to condense on the substrate. The advent of microfluidics has marked a renewed interest in droplet research. In that context, Riegler and co-workers investigated the coalescence of droplets of different fluids. They paid a particular attention to the occurrence of fast or delayed coalescence, \cite{Riegler1,Riegler2}, noting the possible motion of one of the droplets. Lai et al. investigated the mixing potential of coalescing droplets on a surface with a wettability gradient, \cite{Yang10}. Finally, the work of Bico and Qu\'er\'e deserves special mention as it is the closed configuration analogue of the work presented here, \cite{Bico1,Bico2}. The authors observed and analyzed the spontaneous motion of liquid slugs induced by their interaction with a slug of a different fluid. The authors focused on flows confined in capillaries or between parallel plates (Hele-Shaw configuration).

Recently, the coalescence of miscible droplets was investigated theoretically in \cite{sellier_nock_verdier} and it was shown that two important dimensionless parameters govern the actuation mechanism. The first one reflects the magnitude of surface tension gradient which is the driving mechanism and the second one, equivalent to a Peclet number, is representative of the importance of diffusion. Clearly, the numerical results indicated that the larger the surface tension gradient, the faster and the longer the motion. Diffusion, on the other hand, had an adverse effect since it tends to homogenize surface tension and ultimately stop the droplet motion. The work of \cite{sellier_nock_verdier} established the ``theoretical'' existence in the parameter space of a self-propulsion window and the natural follow-up study was to demonstrate experimentally these phenomena. The object of this paper is to report on this experimental work and lay the basis of a simple phenomenological model which attempts to capture some of the underlying physics.

\section{Experimental evidences}
\label{sec:1}
\subsection{Observations on the coalescence of droplets of different fluids}
\begin{figure}[h]
\begin{center}
\resizebox{0.6\columnwidth}{!}{\includegraphics{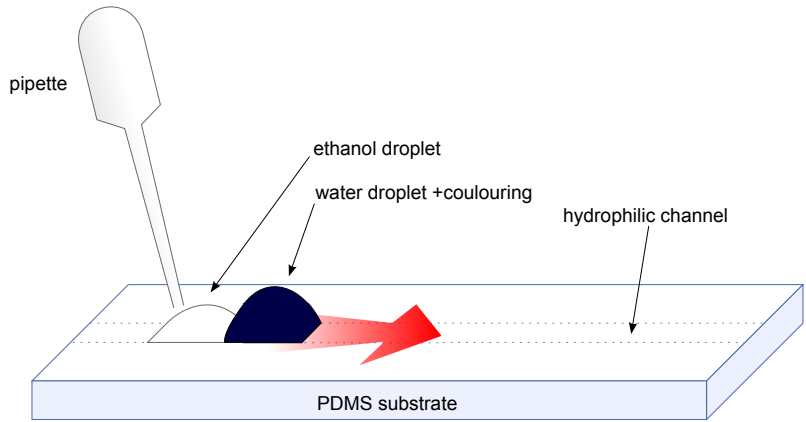}}
\caption{Principle of the experiments: a droplet of water deposited in the middle of an hydrophilic channel is actuated by a droplet of ethanol dropped nearby.}
\label{fig:1}       
\end{center}
\end{figure}

The idea of the experiment is elementary and illustrated in Fig. \ref{fig:1}. A droplet of \textit{Fluid 1} with an approximately known volume was pre-positioned on a substrate using a micro-pipette. A droplet of \textit{Fluid 2} was then deposited with a micro-pipette also in the immediate vicinity of the \textit{Fluid 1} droplet. The flow dynamic was captured using a high-speed camera. The experiments are performed in open, uncontrolled atmospheric conditions on a horizontal substrate. A range of fluid combinations, miscible or immiscible, and substrates were tested.

A wealth of fascinating phenomena was observed during the coalescence. These are highlighted in a video which can be viewed at \url{http://www.youtube.com/watch?v=E33FxyqXLfw}. Some of the most striking features which were observed include:
\begin{itemize}
\item \textbf{Spreading reversal:} for water-ethanol on a Teflon substrate, the initially rapid coalescence and outward spread is followed by a recession of the contact line (inward spread).
\item \textbf{Transition from static to transient states:} for water-ethanol on a PMMA subtrate, the initial coalescence is followed by a relatively long static period during which the droplet does not deform but spreading ultimately resumes reaching a different final equilibrium state.
\item \textbf{Chaotic transport and clustering:} chaotic transport of small particles seeded in a water droplet coalescing with an ethanol droplet. The particles ultimately cluster in the contact line region.
\item \textbf{Contact line fingering and free surface turbulence:} for a toluene-silicone oil system on a glass substrate, the interaction of these two immiscible fluids is very peculiar. The toluene front reacts with the silicone oil droplet literally pushing it away. Fingering instabilities are clearly visible at the toluene contact line.
\end{itemize}
This qualitative description and the range of other combinations we studied allowed us to draw several preliminary observations/conclusions:
\begin{itemize}
\item As expected, droplet actuation is more likely on a highly-wettable substrate with a low apparent contact-angle hysteresis.
\item Actuation over significant distances was only achieved when the actuating fluid was volatile indicating the importance of the vapour phase in the induction of the surface energy gradient. In some of our experiments, the actuated droplet could be displaced by simply approaching (without touching or depositing) the actuating droplet.
\item The motion only lasts a limited amount of time clearly showing that the surface energy gradient is a short lasting phenomenon.
\item In some cases, no apparent thin film connects the actuating and the actuated droplets.
\end{itemize}
It is clear from basic physics considerations that two mechanisms may be responsible for the observed motion. The first one is related to the evaporation of the actuating volatile droplet. The vapour of the volatile droplet diffuses through the surrounding air contaminating the free surface of the other droplet and the substrate. The condensation of the vapour on the substrate results in a one-sided variation of the contact angle of the actuated droplet. This breaks the symmetry of the droplet and creates a pressure gradient which induces propulsion. The local contamination of the free surface by the vapour also creates a surface tension gradient and Marangoni stresses which tend to displace the fluid. The second important mechanism is a consequence of the actual mixing of the two droplet which creates a surface tension gradient. Since the system loses its symmetry during the coalescence, the capillary forces exerted on the contact line does not balance on either side and the bi-drop is ``pulled'' towards the highest surface tensions. The atmosphere plays a passive role in this second mechanism. This mechanism is the one observed by Bico and Qu\'er\'e in confined configurations (bi-slugs in capillaries or between parallel plates), \cite{Bico1,Bico2}.
\subsection{Motion control using substrate patterning}
From a potential microfluidic application perspective, the motion of the droplet needs to be controlled in some sense. The most convenient way to achieve this motion control is to create hydrophilic stripes on the substrate. This was achieved by plasma-treating the exposed surface of a PDMS substrate. PDMS substrates were prepared by mixing Sylgard 184 (Dow Corning) base and curing agent in a 10:1 w/w ratio. The pre-polymer mixture was degassed and poured into round PS culture trays, followed by further degassing. The PDMS was cured 3 h at 65 °C in a convection oven. After removal from the oven, substrates were left 24 h at room temperature before use. Rectangular samples of PDMS were cut out using a scalpel and placed upside down onto standard glass microscope slides for handling. Glass coverslips were used to pattern areas of variable surface energy on the PDMS sample. This was performed by bringing the coverslips in reversible contact with the PDMS and exposing the samples to air plasma for 30 s at 100 W. Areas covered by glass remain hydrophobic, while uncovered areas turn hydrophilic. After plasma treatment, the coverslips were peeled off.

\begin{figure}[h]
\begin{center}
\resizebox{0.9\columnwidth}{!}{\includegraphics{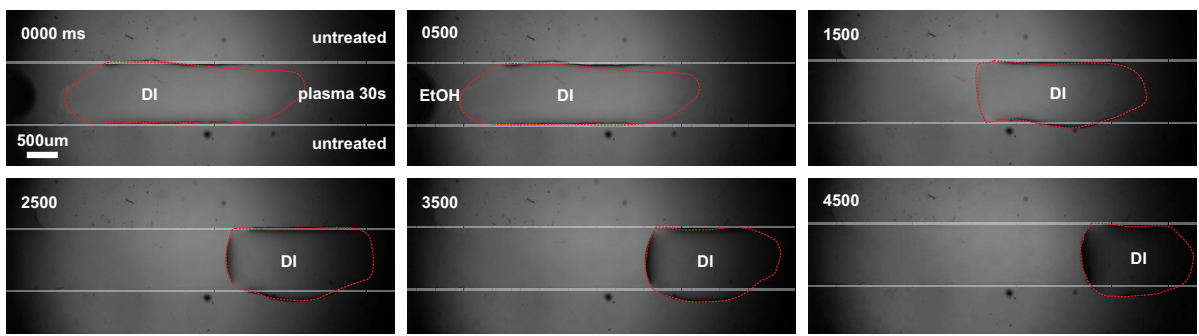}}
\caption{Picture sequence showing from above the actuation of the water droplet by an ethanol droplet on an hydrophilic highway.}
\label{fig:2}       
\end{center}
\end{figure}

The most successful fluid combination on these hydrophilic highways was found to be water and ethanol. Fig. \ref{fig:2} shows a top view of the coalescence on subsequent motion on the stripe. The time sequence reads from top to bottom. The distilled water droplet (labelled DI in Fig. \ref{fig:2}) is first placed on the stripe. The ethanol droplet (labelled EtOH) is deposited to the left of the water droplet (it is barely visible because of the lack of contrast). The water droplet is clearly seen to move to the right-hand side of the stripe. In some cases the motion could last over long distances (up to 10 times the initial droplet diameter).

\begin{figure}[h]
\begin{minipage}[b]{0.5\linewidth}
\centering
\includegraphics[scale=4]{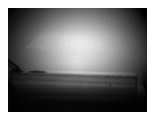}
\end{minipage}
  \begin{minipage}[b]{0.5\linewidth}
\centering
\includegraphics[scale=4]{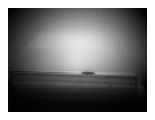}
\end{minipage}
\caption{Side view of the water droplet actuated by the ethanol droplet at two different times. The water droplet moves from left to right.}
\label{fig:3}
\end{figure}

\begin{figure}[h]
\begin{minipage}[b]{0.5\linewidth}
\centering
\includegraphics[scale=0.25]{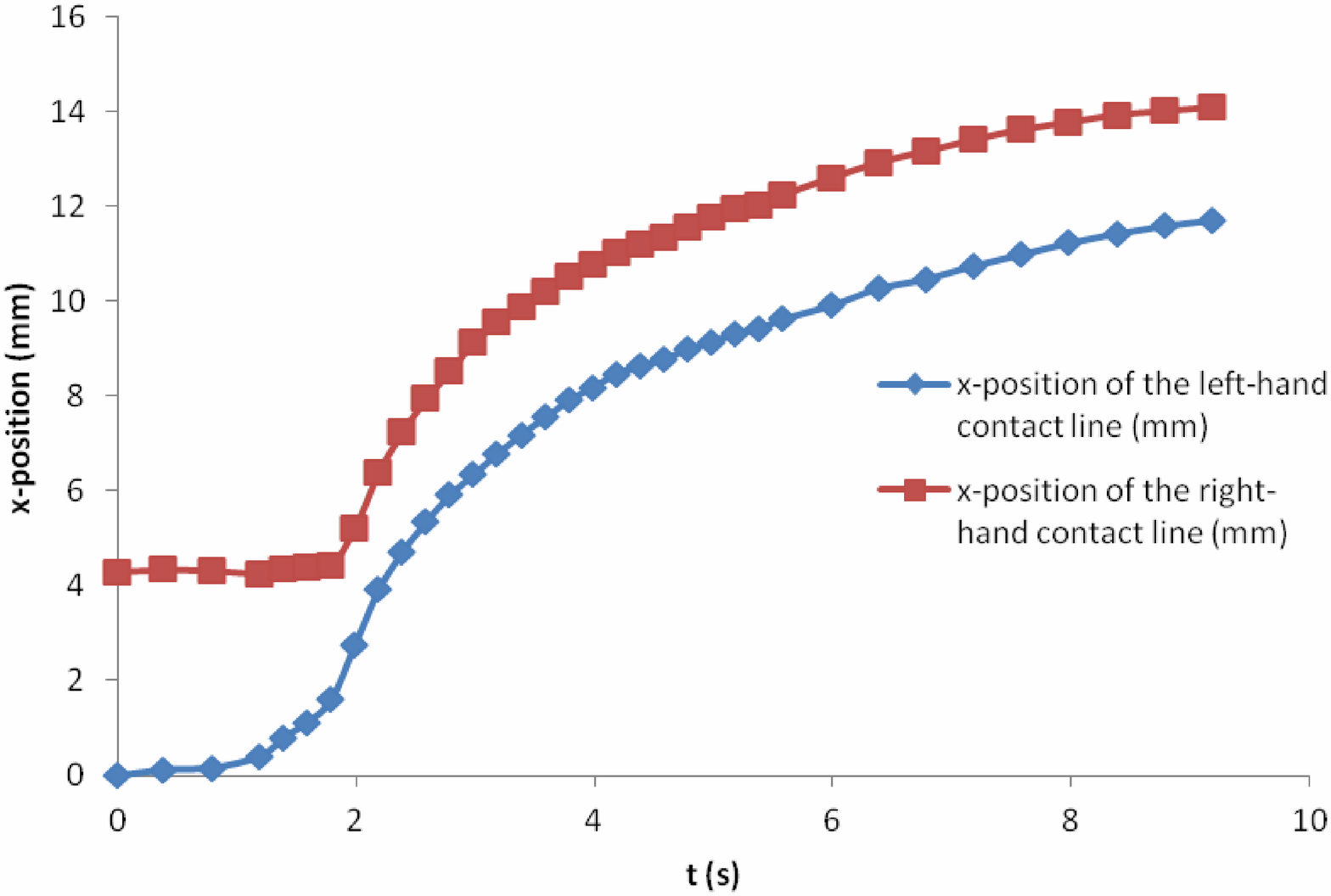}
\end{minipage}
  \begin{minipage}[b]{0.5\linewidth}
\centering
\includegraphics[scale=0.25]{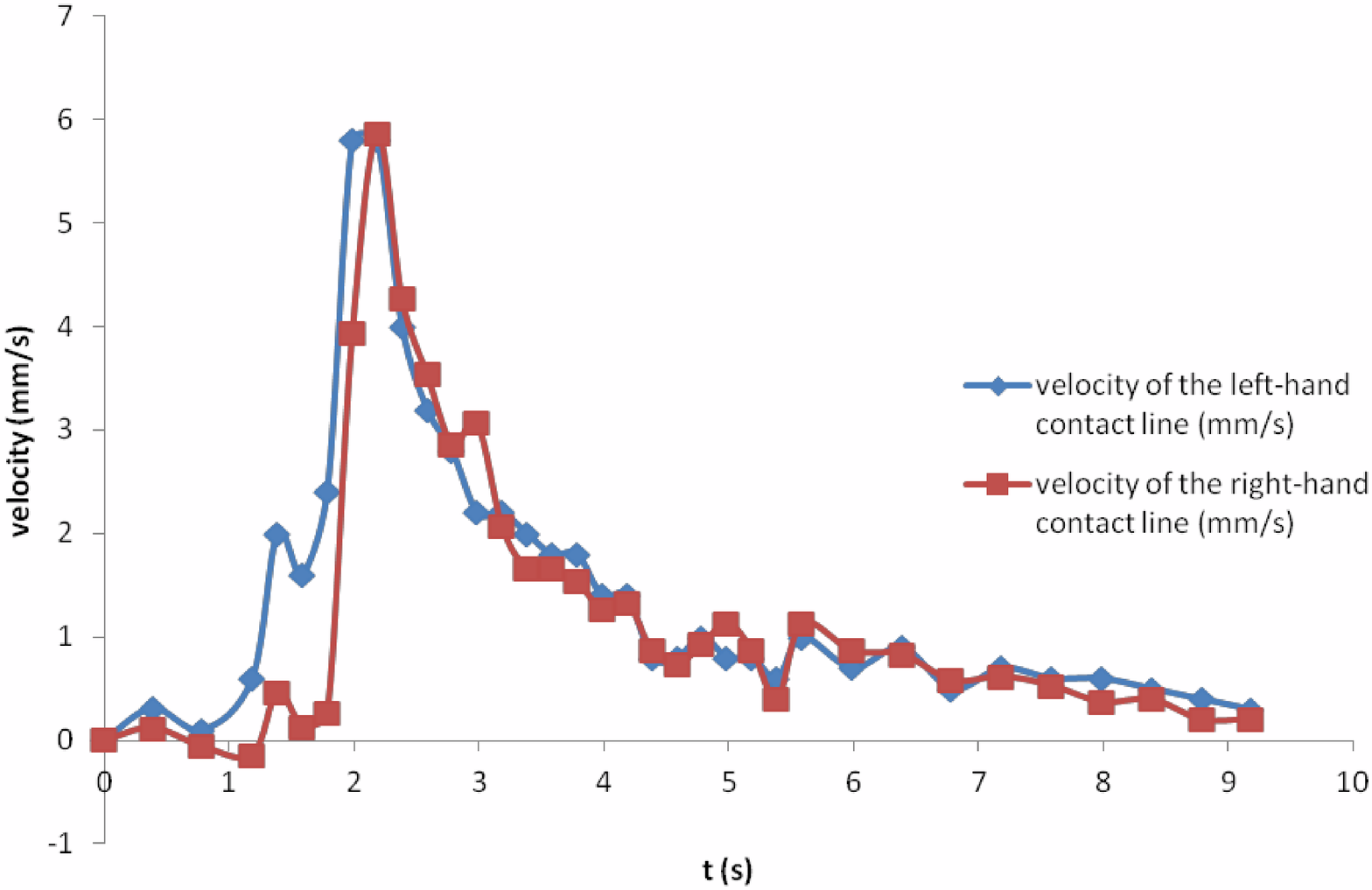}
\end{minipage}
\caption{Position and velocity of the left-hand and right-hand contact lines.}
\label{fig:4}
\end{figure}

The side view of the phenomenon, shown in Fig. \ref{fig:3}, offers greater insight into the dynamics of the droplet. In particular, it is possible to measure the evolution of the advancing and receding contact lines. A side view video of the droplet motion can be seen at \url{http://www.scivee.tv/node/26233}. Images were post-processed using the ImageJ program, \cite{Abramoff04}.
Figure \ref{fig:4} shows the position of the front and rear contact lines as a function of time and the corresponding contact line velocity for the actuated droplet (the water droplet on the right-hand side). A number of interesting features can be observed. The contact line closest to the ethanol (left-hand contact line) starts moving before the right-hand contact line. This results in an initial contraction of the droplet (up to 2~s). The water droplet then starts to move with a near constant diameter (distance from left-hand to right-hand contact lines). The initial motion is quite rapid. The velocity peaks at about 6 mm/s and then decays in an exponential fashion towards zero. This trend was repeatedly observed for several data sets and is in a sense the expectation from \cite{sellier_nock_verdier}.

\begin{figure}[h]
\begin{center}
\resizebox{0.9\columnwidth}{!}{\includegraphics{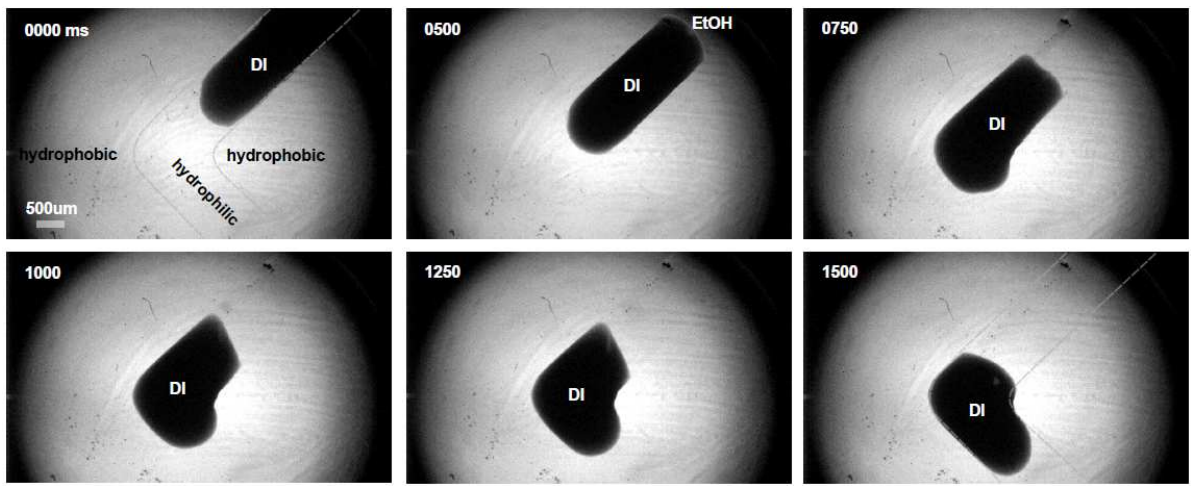}}
\caption{Distilled water droplet forced around a 90$^o$ corner.}
\label{fig:5}       
\end{center}
\end{figure}

The surface treatment enables a control of the actuated droplet trajectory as can be seen in Fig. \ref{fig:5} which illustrates the distilled water droplet going around a 90$^o$ corner after the deposition of an ethanol droplet at the top of the pictures.
The prospect to manipulate droplets on patterned surfaces opens up interesting opportunities in the field of microfluidics. The following section presents an initial modeling attempt.
\section{Phenomenological model}
\begin{figure}
\begin{center}
\resizebox{0.6\columnwidth}{!}{\includegraphics{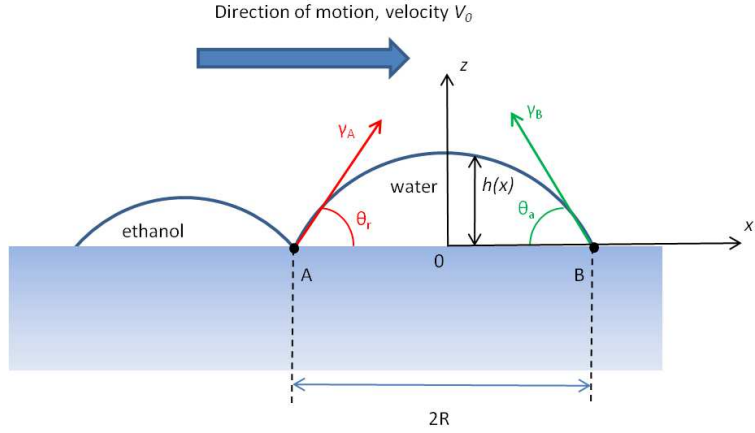}}
\caption{Sketch of the droplet system and notations.}
\label{fig:6}       
\end{center}
\end{figure}

This section describes a simple model capturing the velocity variation shown in Fig. \ref{fig:4}. In order to shed some light on the underlying physics, we consider the simpler quasi-two-dimensional system shown in Fig. \ref{fig:6} which consists of a water droplet interacting with an ethanol droplet and follow an analysis analogous to the one presented in \cite{deGennes_book,Ford94} for thermocapillary actuation.
The proposed scenario is that, as the ethanol droplet approaches the water droplet, it modifies the local surface energies and induces the water droplet motion. Experimentally, it is observed that the water droplet shape does not vary much once it has been actuated and therefore, it will be assumed that the droplet profile remains constant in the following. The motion is the results of a balance between the surface energy driving force, $F_d$, and the viscous resistive force, $F_r$. Accordingly, Newton's second law requires that
\begin{equation}
m\frac{dV}{dt} = F_d - F_r \; ,
\label{eq:1}
\end{equation}
where $V$ is the actuated droplet velocity and $m$ its mass. The driving force on the water droplet can be expressed as
\begin{equation}
F_d = w \left( \gamma _B \cos \theta _a - \gamma _A \cos \theta _r \right) \; ,
\label{eq:2}
\end{equation}
where $w$ is the width of the hydrophilic stripe, $\theta _a$ and $\theta _r$ the advancing and receding contact angles for the water droplet (on the right-hand side and left-hand side in Fig. \ref{fig:6}, respectively) and $\gamma _B$ and $\gamma _A$ the respective surface tensions there. The resistive force is estimated by calculating the shear force exerted by the substrate on the droplet. The flow field is best inferred in a reference frame centered at the droplet centre which moves with velocity $V$. Assuming a unidirectional flow on the hydrophilic stripe, the lubrication approximation enables the evaluation of the velocity profile. Accordingly,
\begin{equation}
\mu u(z) = \frac{1}{2} \frac{dp}{dx} \left( z^2-2zh \right) + \frac{d\gamma}{dx}z - \mu V \; ,
\label{eq:3}
\end{equation}
where $\mu$ is the dynamic viscosity, $\frac{dp}{dx}$ the pressure gradient yet to be determined, and $\frac{d\gamma}{dx}$ the surface tension gradient. This velocity profile satisfies the continuity of tangential stress at the free surface and the no-slip condition at the substrate. The pressure gradient in the velocity profile can be inferred by invoking the conservation of mass. After some algebraic manipulation, the wall shear stress $\tau_w$ is found to be given by
\begin{equation}
\tau _w = \frac{3\mu V}{h}-\frac{1}{2} \frac{d\gamma}{dx} \; .
\label{eq:4}
\end{equation}
Finally, the resistive force is found by integrating the wall shear stress over the whole footprint of the droplet,
\begin{equation}
F_r = \int _{-R+\epsilon}^{R-\epsilon} \int _{-\frac{w}{2}+\epsilon}^{\frac{w}{2}-\epsilon} \tau _w dxdy \; .
\label{eq:5}
\end{equation}
Following \cite{Subramanian05,Pratap08}, we introduce the cut-off length $\epsilon$ to circumvent the stress singularity at the contact line. This length is representative of the region where the no-slip condition is expected to break down and is therefore expected to be on the order of molecular dimensions. At this stage, some assumptions are necessary regarding the droplet profile. Because the width of the hydrophilic channel is smaller than the characteristic Capillary length, it is safe to assume that the cross-flow droplet profile will be parabolic and following Darhuber and co-workers, \cite{Darhuber03}, we express the droplet thickness profile as
\begin{equation}
h(x,y) = h_c(x) \left( 1-4\frac{y^2}{w^2} \right) \; ,
\label{eq:6}
\end{equation}
where $h_c (x)$ is the droplet centreline profile. Substituting eq. (\ref{eq:4}) into eq. (\ref{eq:5}) along with the assumed droplet shape (eq. (\ref{eq:6}) leads to
\begin{equation}
F_r = -\frac{w}{2} \left( \gamma _B - \gamma _A \right) + \frac{w}{2} \log \left| \frac{w-\epsilon}{\epsilon} \right| 3 \mu V \int _{-R+\epsilon}^{R-\epsilon} \frac{1}{h_c}dx \; .
\label{eq:7}
\end{equation}
Moreover, following Ford and Nadim in \cite{Ford94}, we assume the droplet to have approximately a parabolic profile such that
\begin{equation}
h_c(x) = \frac{\tan \theta _s}{2R} \left( R^2-x^2 \right) \; ,
\label{eq:8}
\end{equation}
where $\theta _s$ is the static contact angle. With this final assumption regarding the droplet profile, eq. (\ref{eq:7}) may finally be written as
\begin{equation}
F_r = -\frac{w}{2} \left( \gamma _B - \gamma _A \right) + \log \left| \frac{w-\epsilon}{\epsilon} \right| \log \left| \frac{2R-\epsilon}{\epsilon} \right| \frac{3w\mu V}{\tan \theta _s} \; .
\label{eq:9}
\end{equation}
Finally, combining eqs. (\ref{eq:1}), (\ref{eq:2}), and (\ref{eq:9}), we obtain
\begin{equation}
m\frac{dV}{dt} = w \left( \gamma _B \cos \theta _a - \gamma _A \cos \theta _r \right)+\frac{w}{2} \left( \gamma _B - \gamma _A \right) - \log \left| \frac{w-\epsilon}{\epsilon} \right| \log \left| \frac{2R-\epsilon}{\epsilon} \right| \frac{3w\mu V}{\tan \theta _s} \; .
\label{eq:10}
\end{equation}
As the velocity peaks for $\frac{dV}{dt}=0$, this condition allows us to estimate the peak velocity $V_{max}$ which is given by
\begin{equation}
V_{max} = \frac{\tan \theta _s}{3\mu K} \left( \left( \gamma _B^\star \cos \theta _a - \gamma _A^\star \cos \theta _r \right)+\frac{1}{2} \left( \gamma _B^\star - \gamma _A^\star \right) \right), \textrm{with} \; K=\log \left| \frac{w-\epsilon}{\epsilon} \right| \log \left| \frac{2R-\epsilon}{\epsilon} \right|,
\label{eq:Vmax}
\end{equation}
where the $\star$ indicates surface tension values evaluated at the time when the velocity peaks.
If we assume that $\theta _a = \theta _s - \Delta \theta$ and $\theta _r = \theta _s + \Delta \theta$ where $\Delta \theta$ is small, we can also rewrite the peak velocity as follows:
\begin{equation}
V_{max} = \frac{\tan \theta _s}{3\mu K} \left( \left( \gamma _B^\star - \gamma _A^\star \right) \left( \cos \theta _s +\frac{1}{2} \right) + \left( \gamma _B^\star + \gamma _A^\star \right) \sin \theta _s \Delta \theta \right) \; .
\label{eq:Vmax2}
\end{equation}
From this expression, we conclude that the peak velocity increases with the surface tension difference $\left( \gamma _B - \gamma _A \right)$. This conclusion was also predicted in \cite{sellier_nock_verdier}. The peak velocity also increases with the static contact angle, and the contact angle difference $\Delta \theta$. On the other hand, the peak velocity decreases with the viscosity as intuitively expected. Interestingly, it can be noted that this peak velocity appears to be independent of the width of the hydrophilic channel. The reason for this is that both forces vary linearly with the width of the channel. Note that this observation was confirmed experimentally as no discernable effect of the channel width on the distance traveled was observed. 

With the assumed droplet profile, the mass can easily be expressed in terms of observable quantities according to
\begin{equation}
m = \rho \frac{8}{18} w \tan \theta _s R^2 \; .
\label{eq:mass}
\end{equation}
Equation (\ref{eq:10}) is a simple ordinary differential equation which can easily be integrated but the question now arises as to how to compute $\gamma _A$ and $\gamma _B$. As a first approximation, we make the assumption that the ethanol vapour ``dissolves'' in water on the ethanol side thereby decreasing the surface tension. Only a limited amount $c_{sat}$ of ethanol is able to dissolve. In the proposed scenario, the ethanol vapour diffuses through the surrounding atmosphere with a diffusion constant $D$ and progressively contaminates the water droplet free surface thereby decreasing its surface tension. We assume this diffusion process to be one-dimensional as a first approximation (i.e. uniform ethanol concentration across the droplet cross-section). Accordingly, the ethanol concentration satisfies
\begin{eqnarray}
& & \frac{\partial c}{\partial t} = \frac{\partial}{\partial x} \left( D \frac{\partial c}{\partial x} \right) \; , \\
& & c(x,t=0) = 0 \; , \\
& & c(x=x_A,t>0) = c_{sat} \; , \\
& & \frac{\partial c}{\partial x} | _{x_b} = 0 \; .
\end{eqnarray}
The left-hand boundary condition implies that on the ethanol side, the ethanol concentration is equal to $c_{sat}$ and the right-hand condition is essentially a no-flux boundary condition. This diffusion equation admits the following analytical solution, \cite{Kreyszig}:
\begin{equation}
c(x,t) = c_{sat}-c_{sat} \sum _{n=0}^{\infty} B_n \sin \left( \left( 2n+1 \right) \frac{\pi x}{4R} \right) \exp ^{-\frac{(2n+1)^2\pi ^2 Dt}{16R^2}} \; , \textrm{with} \; B_n = \frac{4}{(2n+1)\pi} \; .
\label{eq:11}
\end{equation}
surface tension of this water-ethanol mixture is reported in \cite{Vasquez95} and well approximated by
\begin{equation}
\gamma _{mix}(c) = 0.023+0.047\exp ^{-\frac{c}{0.1713}} \; .
\label{eq:12}
\end{equation}
Using eqs. (\ref{eq:10}), (\ref{eq:11}), and (\ref{eq:12}), the variation of the droplet velocity with time can be computed for given values of the advancing and receding contact angles. These are taken to be time-averaged values over the picture sequence extracted from the recorded video (see Fig. \ref{fig:3} for two such pictures). Accordingly, the advancing contact angle $\theta _a$ is taken as 34$^o$ and the receding one $\theta _r$ as 35$^o$. The static contact angle is chosen to be the average of the advancing and receding ones. The droplet radius, also averaged over time, is 1.3 mm. The dynamic viscosity is chosen to be that of water, i.e. 10$^{-3}$ Pa.s.
This model has three parameters for which only limited knowledge is available:
\begin{itemize}
\item The cut-off length $\epsilon$. This cut-off length is representative of a molecular length scale. We choose here a value of $10^{-9}$m which is comparable to the values chosen in \cite{Subramanian05}.
\item The diffusion constant $D$. Reasonable values of diffusion constants for the vapour of various volatile liquids have been reported in \cite{Sultan05} and many typical values are close to $3\times 10^{-6}$ m$^2$/s. We therefore choose here this particular value. Note that diffusion in the liquid bulk is order of magnitude slower.
\item The {\it saturation concentration} $c_{sat}$. This introduced parameter is the least quantifiable one and it will serve as a fitting parameter.
\end{itemize}

\begin{figure}[h]
\begin{minipage}[b]{0.5\linewidth}
\centering
\includegraphics[scale=0.6]{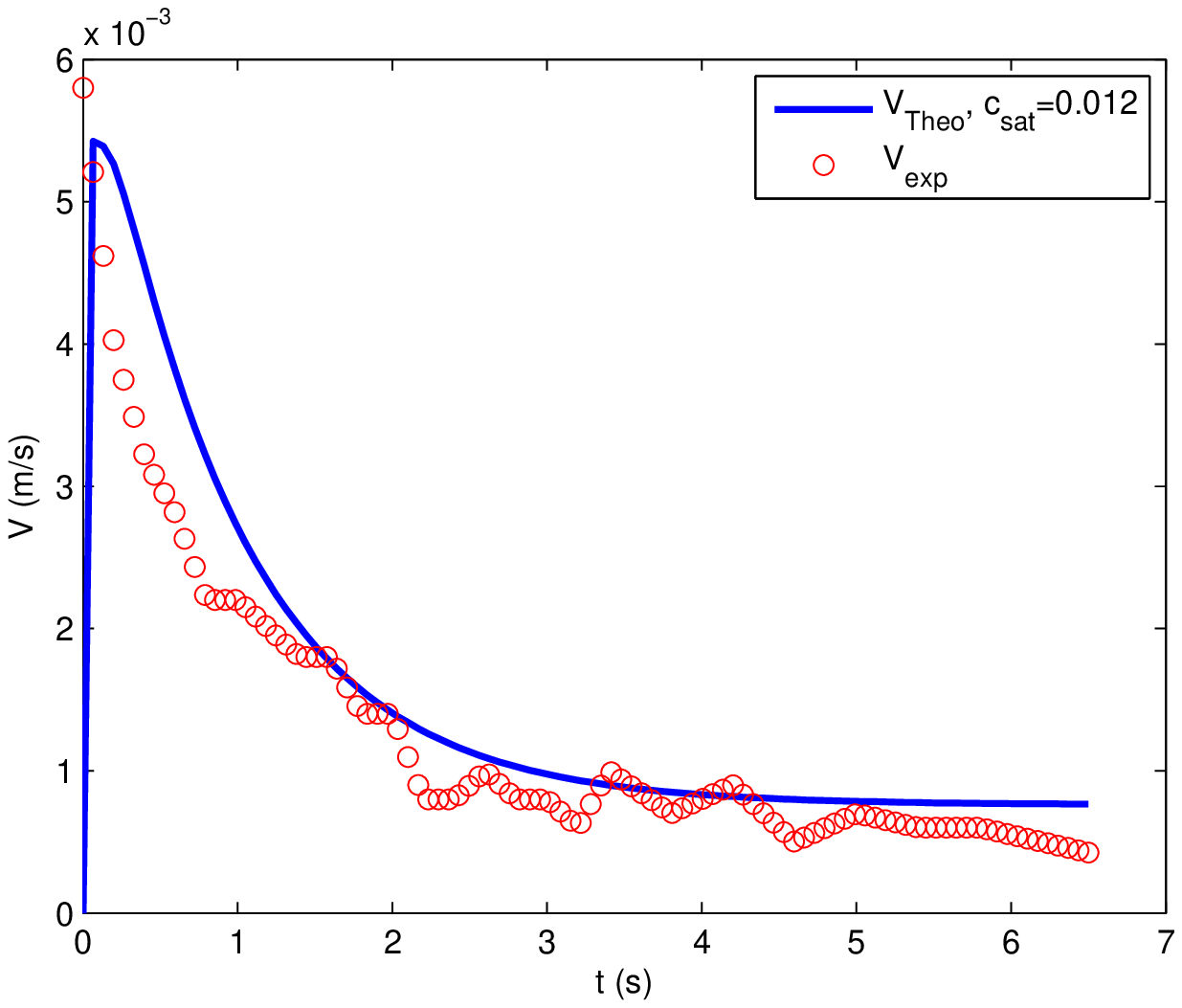}
\end{minipage}
  \begin{minipage}[b]{0.5\linewidth}
\centering
\includegraphics[scale=0.6]{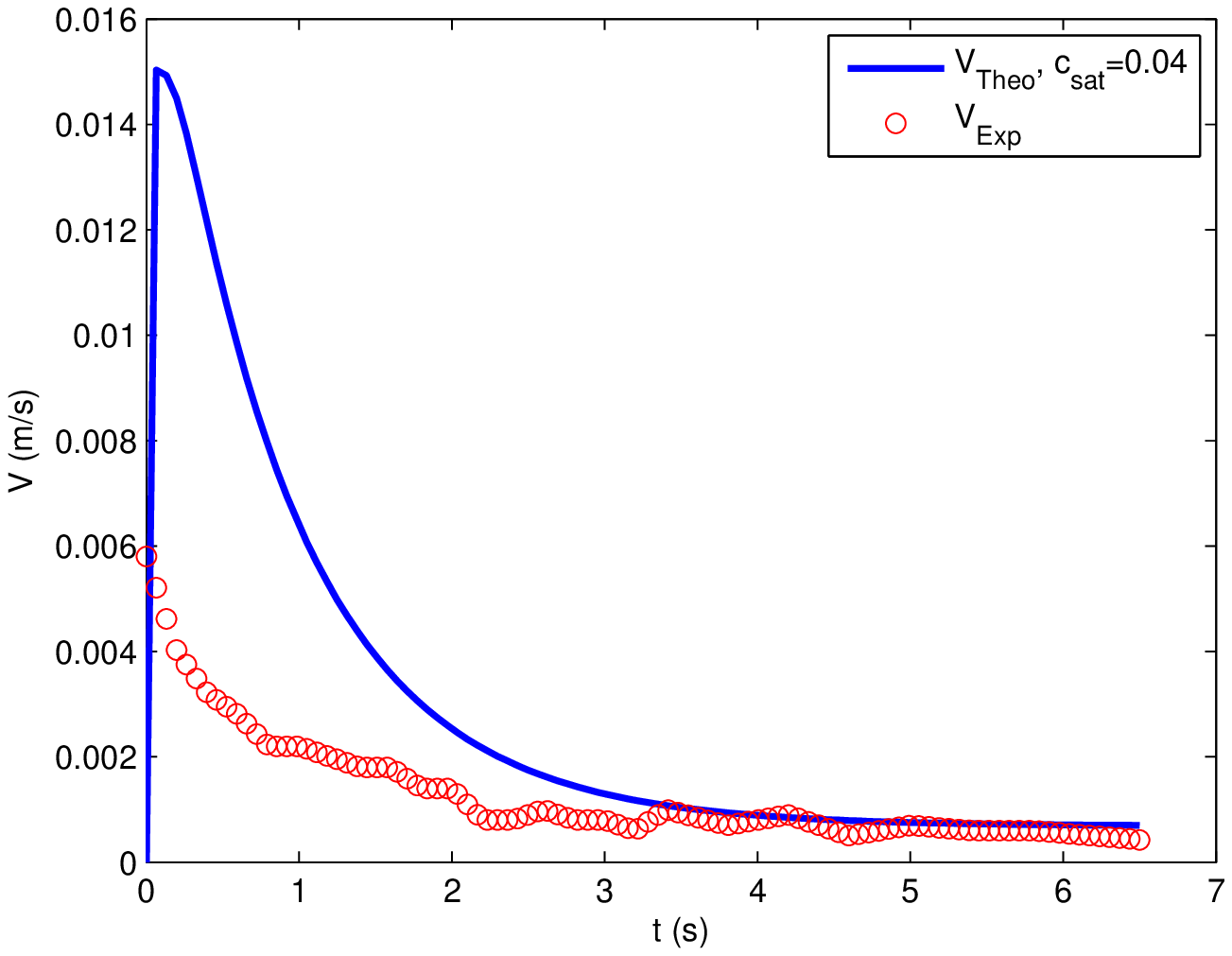}
\end{minipage}
\caption{Comparison of the experimentally measured droplet velocity (red dots) and the theoretically predicted one (blue line). The left hand figure is for $c_{sat}$  = 0.012, and the right-hand one for $c_{sat}$ = 0.04.}
\label{fig:7}
\end{figure}
The differential equation (\ref{eq:10}) is simply solved using the {\it ode45} MATLAB routine. Results reported in Fig. \ref{fig:7} are only for the exponential decay which follows the velocity peaks observed in Fig. \ref{fig:4}. The left-hand graph with $c_{sat} = 0.012$ reproduces very closely the experimentally observed velocity peak and the velocity decay. The peak velocity is most influenced by the value of $c_{sat}$ as larger values of this parameter result in a larger surface tension difference between both ends of the droplets and therefore a greater driving potential. This is observed on the right-hand graph of Fig. \ref{fig:7} which shows that the peak velocity increases by a factor of two as $c_{sat}$ is increased from 0.012 to 0.04. It should be noted that this low value of $c_{sat}$ implies that a relatively small surface tension gradient is actually present at the water droplet surface. This surface tension gradient is on the order of 1.2$\times 10^{-3}$ N/m/mm. The diffusion constant on the other hand has most effect on the time constant in the velocity decay as intuitively expected.
\section{Conclusions}
This paper discusses an actuation mechanism with potential application in the field of microfluidics. This mechanism relies on the surface energy gradient which arises when two droplets are placed in immediate proximity and coalesce. We have studied a wide range of fluid systems and observed complex phenomena including spreading reversal, transitions from static to transient states, chaotic particle transport, particle clustering at the contact lines, contact line fingering, and free surface turbulence. The most reactive fluid combinations always involved a volatile fluid phase indicating the importance of the participating atmosphere. With microfluidics application in mind, we also presented a possible way to control the induced motion of one of the droplets by creating a hydrophilic ``highway'' on a PDMS substrate. We have shown that this is a convenient way to control the droplet motion in straight line but also on more complex paths such as a corner. Concentrating on the simplest case of a water droplet actuated by an ethanol droplet on a straight hydrophilic channel, we observed that the water droplet velocity rapidly reaches a maximum value of around half a centimeter per second and then slowly decays in an apparent exponential fashion over a period of around 6 seconds. Note that although only one case was presented in detail, this trend was repeatedly observed. In order to explain the underlying physics, we have developed a simple model based on Newton's second law applied to the water droplet and taking into account the driving force resulting from gradients of surface energies and the viscous resistive force. In this simple model, one unknown parameter, $c_{sat}$, served as an adjustable parameter that allowed the model to closely capture the dynamics of the water droplet and predicts the effect of various parameters such as the viscosity, the contact angles, or the surface tension gradient on the peak velocity. The results of the model suggest that the decrease of surface tension of the water droplet on the ethanol side is quite small, on the order of 0.0032 N/m.
\section*{Acknowledgements}
The authors gratefully acknowledge the support of the Royal Society of New Zealand through a Marsden Grant (Grant number UOC1104).

\end{document}